\documentclass{amsart}
\usepackage{amsfonts}

\usepackage{graphicx}

\input{tcilatex}
\input tcilatex

\begin{document}
\title[Moran Model with Mutations]{Karlin-McGregor mutational occupancy problem revisited}
\author{Thierry E. Huillet}
\address{Laboratoire de Physique Th\'{e}orique et Mod\'{e}lisation \\
CNRS-UMR 8089 et Universit\'{e} de Cergy-Pontoise, 2 Avenue Adolphe Chauvin,
95302, Cergy-Pontoise, FRANCE\\
E-mail: Thierry.Huillet@u-cergy.fr }
\maketitle

\begin{abstract}
Some population is made of $n$ individuals that can be of $p$ possible
species (or types). The update of the species abundance occupancies is from
a Moran mutational model designed by Karlin and McGregor in 1967. We first
study the equilibrium species counts as a function of $n,$ $p$ and the total
mutation probability $\nu $ before considering various asymptotic regimes on 
$n$, $p$ and $\nu .$\newline

\textbf{Running title:} KMG Model with Mutations.\newline

\newline
\end{abstract}

\section{Introduction}

Some population is made of $n$ individuals that can be of $p$ possible
species (or types). The discrete-time update of the species abundance
occupancies is from a Moran mutational model first designed in \cite{KMG}
and for which the size $n$ of the population is maintained constant over the
generations. We will study in great detail the equilibrium species counts as
a function of $n,$ $p$ and the total mutation probability $\nu $ before
considering various asymptotic regimes of interest on $n$, $p$ and $\nu $,
some of which were not considered in \cite{KMG}. When they exist while $%
n\wedge p\rightarrow \infty $, the limiting distributions of the typical
species abundance are not heavy-tailed, rather they have a dominant
exponential decay factor and this may be seen to result from the
conservation of the population size $n$. They are rather related to the
negative binomial or Fisher log-series distributions, \cite{FCW}. Also of
particular interest will be $\left( i\right) $ the distribution of the
number of occupied species with positive occupancy $\left( ii\right) $ the
probability that two randomly sampled individuals are of the same species;
this both for fixed $n$, $p$ and $\nu $ and under their asymptotic regimes.

This model should not be confused with the following related (although
non-conservative) Yule mutation model, \cite{WY}, \cite{Y}: A species starts
with a single individual. As a result of mutations, new individuals are
produced according to a linear pure birth Yule process with some birth rate
and they all belong to the same species. Concomitantly and as a result of
specific mutations, inside a species, an individual of a novel species can
be created at some other rate and the new species, once it has appeared,
behaves like all the previous ones. For the Yule model, the asymptotic
abundance inside a typical species is distributed like a Simon distribution 
\cite{S} which (in sharp contrast with the former log-series-like
distribution), is heavy-tailed, translating the presence of very large
family counts. Note that here both $n$ and the number $p$ of possible
species should be set to infinity because both are bound to grow
indefinitely in the process, see \cite{SR}.

The Karlin-McGregor (KMG) mutation model was originally developed to study
multiallelic frequencies dynamics in population genetics, as from \cite{KC}.
It was later applied to the study of surname distributions and random
isonymy, making the observation that surnames can be considered as alleles
transmitted along the male line. See \cite{YCS}, \cite{MZ}, \cite{Zei}, \cite
{R} and the references therein. One can apply the model not only to surnames
(which can be linked to $Y-$chromosomes) but also to first names and other
elements of culture that do propagate by copying.

\section{Species abundances evolution: the KMG mutation model}

Some population is made of $n$ individuals that can be of $p$ possible
species (or types).

At (discrete-time) step $t$, there are $K_{t}\left( q\right) \geq 0$
individuals of type $q$, $q=1,...,p.$ The occupancy vector $\mathbf{K}%
_{t}:=\left( K_{t}\left( q\right) ;q=1,...,p\right) $ is called the species
abundance vector. The species $q$ will be said filled if $K_{t}\left(
q\right) >0$ (it has at least one representative).

We let $Q_{t}:=\sum_{q=1}^{p}\mathbf{1}\left( K_{t}\left( q\right) >0\right) 
$ be the number of types present at step $t$ (the number of filled species).

We let $N_{t}\left( k\right) :=\sum_{q=1}^{p}\mathbf{1}\left( K_{t}\left(
q\right) =k\right) $ be the number of species with $k$ representatives at
step $t$.

We have $1\leq Q_{t}=p-N_{t}\left( 0\right) \leq $ $p\wedge n$ and $%
\sum_{q=1}^{p}K_{t}\left( q\right) =n=\sum_{k=1}^{\max_{q}K_{t}\left(
q\right) }kN_{t}\left( k\right) .$

The dynamics of $\mathbf{K}_{t}$ is in the spirit of a Moran $\beta -$%
mutation evolution model, preserving the total number of individuals $n,$
namely, \cite{Moran}, \cite{Hui}:

Given $K_{t}\left( q\right) =k_{q},$ $q=1,...,p$, we let $\left(
k_{1},...,k_{p}\right) \rightarrow \left( k_{1},...,k_{q}-1,...,k_{q^{\prime
}}+1,...,k_{p}\right) $ be the moves between step $t$ and step $t+1$: at
each step, an individual of type $q$ is deleted from the population and an
individual of type $q^{\prime }\neq q$ is created. We assume that this event
occurs with probability (w.p.) 
\begin{equation}
\frac{k_{q}}{n}\left[ \frac{k_{q^{\prime }}}{n}\left( 1-\left( p-1\right)
\beta \right) +\left( 1-\frac{k_{q^{\prime }}}{n}\right) \beta \right] .
\label{f1}
\end{equation}
For such a mutation model, an individual of type $q$ is deleted (with
probability $\frac{k_{q}}{n}$) and an individual of type $q^{\prime }$ is
created either because $q^{\prime }$ is selected to duplicate (with
probability $\frac{k_{q^{\prime }}}{n}$) and the duplicate has not mutated
to any other state than $q^{\prime }$ (an event of probability $1-\left(
p-1\right) \beta $) or because an individual of type $q^{\prime \prime }\neq
q^{\prime }$ is selected to duplicate (with probability $1-\frac{%
k_{q^{\prime }}}{n}$) and the duplicate has mutated to an individual of type 
$q^{\prime }$ (with probability $\beta $). We let $p\beta =\nu $ be the
overall mutation probability.

When the $K_{t=0}\left( q\right) $'s are exchangeable, the $K_{t}\left(
q\right) $'s remain exchangeable for all $t$ (having law invariant upon a
permutation of the $q$'s), in particular all the $K_{t}\left( q\right) $'s
share the same distribution. Let us thus focus on $K_{t}\left( 1\right) $
with $K_{t}\left( q\right) \overset{d}{=}K_{t}\left( 1\right) $, $q=2,...,p$
(equality in distribution). Then, see \cite{KMG}, while lumping the states $%
K_{t}\left( q\right) $, $q=2,...,n$, given $K_{t}\left( 1\right) =k\in
\left\{ 0,...,n\right\} $%
\begin{equation*}
\begin{array}{l}
\left( k,n-k\right) \rightarrow \left( k+1,n-k-1\right) \text{ w.p. }%
p_{k}=\left( 1-\frac{k}{n}\right) \left( \frac{k}{n}\left( 1-\left(
p-1\right) \beta \right) +\left( 1-\frac{k}{n}\right) \beta \right) \\ 
\left( k,n-k\right) \rightarrow \left( k-1,n-k+1\right) \text{ w.p. }q_{k}=%
\frac{k}{n}\left( \frac{k}{n}\left( p-1\right) \beta +\left( 1-\frac{k}{n}%
\right) \left( 1-\beta \right) \right)
\end{array}
\end{equation*}
defines the tridiagonal transition probabilities of a random walk on the set 
$\left\{ 0,...,n\right\} $ with holding probability $r_{k}=1-\left(
p_{k}+q_{k}\right) $ that $\left( k,n-k\right) \rightarrow \left(
k,n-k\right) .$ This random walk is ergodic with invariant probability
measure (independent of the initial condition $K_{t=0}\left( 1\right) $)
given for $k=0,...,n$ by (see \cite{KMG2} or \cite{Hui} for instance): 
\begin{equation}
\pi _{k}:=\mathbf{P}\left( K_{\infty }\left( 1\right) =k\right) =\frac{%
\binom{k+\frac{1}{p}n\nu /\left( 1-\nu \right) -1}{k}\binom{n/\left( 1-\nu
\right) -k-\frac{1}{p}n\nu /\left( 1-\nu \right) -1}{n-k}}{\binom{n/\left(
1-\nu \right) -1}{n}}.  \label{f2}
\end{equation}
This is also 
\begin{equation}
\begin{array}{l}
\pi _{k}=\binom{n}{k}\frac{B\left( k+\theta ,n-k+\left( p-1\right) \theta
\right) }{B\left( \theta ,\left( p-1\right) \theta \right) } \\ 
=\binom{n}{k}\frac{\Gamma \left( n\nu /\left( 1-\nu \right) \right) }{\Gamma
\left( n/\left( 1-\nu \right) \right) }\frac{\Gamma \left( k+\theta \right) 
}{\Gamma \left( \theta \right) }\frac{\Gamma \left( n/\left( 1-\nu \right)
-k-\theta \right) }{\Gamma \left( n\nu /\left( 1-\nu \right) -\theta \right) 
}
\end{array}
\label{f3}
\end{equation}
where $\theta =\frac{n}{p}\nu /\left( 1-\nu \right) $ and $B\left(
a,b\right) $ is the beta function. In particular, $\pi _{0}=\frac{\Gamma
\left( n\nu /\left( 1-\nu \right) \right) }{\Gamma \left( n/\left( 1-\nu
\right) \right) }\frac{\Gamma \left( n/\left( 1-\nu \right) -\theta \right) 
}{\Gamma \left( n\nu /\left( 1-\nu \right) -\theta \right) }$. The
distribution $\pi _{k}$ of $K\left( 1\right) :=K_{\infty }\left( 1\right) $
is a $B\left( \theta ,\left( p-1\right) \theta \right) $ $s$-mixture of a
binomial bin$\left( n,s\right) $ distribution, $s\in \left( 0,1\right) $. It
is a P\'{o}lya-Eggenberger distribution with probability generating function
(pgf) 
\begin{equation*}
\mathbf{E}\left( u^{K\left( 1\right) }\right) =F\left( -n,\theta ;p\theta
;1-u\right) ,
\end{equation*}
where $F:=_{2}F_{1}$ is a Gauss hypergeometric function. One can check that $%
K\left( 1\right) $ has mean $\mathbf{E}\left( K\left( 1\right) \right) =n/p$
and variance 
\begin{equation*}
\sigma ^{2}\left( K\left( 1\right) \right) =\frac{n\left( p-1\right) \left(
n+p\theta \right) }{p^{2}\left( p\theta +1\right) }=\left( \frac{n}{p}%
\right) ^{2}\frac{p-1}{1+\nu \left( n-1\right) }.
\end{equation*}
An interesting immediate consequence is the following: noting that $%
p_{k}:=k\pi _{k}/\mathbf{E}\left( K\left( 1\right) \right) $ is the
size-biased probability to pick an individual with $k$ representatives at
equilibrium, the probability $\alpha $ that two randomly chosen individuals
from the population are of the same species is 
\begin{equation}
\alpha =\sum_{k=1}^{n}\frac{k}{n}p_{k}=\frac{p}{n^{2}}\sum_{k=1}^{n}k^{2}\pi
_{k}=\frac{p}{n^{2}}\left( \sigma ^{2}\left( K\left( 1\right) \right) +%
\mathbf{E}\left( K\left( 1\right) \right) ^{2}\right) =\frac{p+\nu \left(
n-1\right) }{p\left( 1+\nu \left( n-1\right) \right) }.  \label{f3a}
\end{equation}

The one-dimensional law of $K\left( 1\right) $ being under control for all $%
n,p$, we now wish to evaluate its asymptotic shape under various limiting
conditions on $n,p$, namely $n\approx p$, $n\ll p$ and $n\gg p$
corresponding respectively to $\mu :=n/p=O\left( 1\right) $, $\mu
\rightarrow 0$ and $\mu \rightarrow \infty .$ For each asymptotic regime, we
shall denote by ``$^{*}$'' the asymptotic evaluation of the quantities of
interest.

\section{Various asymptotics}

We shall study five asymptotic regimes depending on the density $\mu $ of
individuals over the species range.

\textbf{1. }(balanced case)\textbf{.} If both $p,n\rightarrow \infty $ while 
$\mu =n/p\rightarrow \mu ^{*}>0$ and $\nu $ fixed, then $\theta =\frac{n}{p}%
\nu /\left( 1-\nu \right) \sim \theta ^{*}=\mu ^{*}\nu /\left( 1-\nu \right)
>0$ and 
\begin{equation}
\begin{array}{l}
\pi _{k}=\binom{n}{k}\frac{\Gamma \left( n\nu /\left( 1-\nu \right) \right) 
}{\Gamma \left( n/\left( 1-\nu \right) \right) }\frac{\Gamma \left( k+\theta
\right) }{\Gamma \left( \theta \right) }\frac{\Gamma \left( n/\left( 1-\nu
\right) -k-\theta \right) }{\Gamma \left( n\nu /\left( 1-\nu \right) -\theta
\right) } \\ 
\sim \frac{n^{k}}{k!}\frac{\Gamma \left( k+\theta ^{*}\right) }{\Gamma
\left( \theta ^{*}\right) }\frac{\left( n/\left( 1-\nu \right) \right)
^{-\left( k+\theta ^{*}\right) }}{\left( n\nu /\left( 1-\nu \right) \right)
^{-\theta ^{*}}}=\frac{\nu ^{\theta ^{*}}}{k!}\frac{\Gamma \left( k+\theta
^{*}\right) }{\Gamma \left( \theta ^{*}\right) }\left( 1-\nu \right)
^{k}=:\pi _{k}^{*}
\end{array}
\label{f4}
\end{equation}
is a well-defined negative binomial distribution for all $k\in \left\{
0,1,2,...\right\} $, so with limiting pgf $\mathbf{E}\left( u^{K\left(
1\right) }\right) \sim \left( \nu /\left( 1-\left( 1-\nu \right) u\right)
\right) ^{\theta ^{*}}$. When $k$ is large $\pi _{k}^{*}\sim \frac{\nu
^{\theta ^{*}}}{k!}\frac{\Gamma \left( k+\theta ^{*}\right) }{\Gamma \left(
\theta ^{*}\right) }\left( 1-\nu \right) ^{k}\sim \frac{\nu ^{\theta ^{*}}}{%
\Gamma \left( \theta ^{*}\right) }k^{\theta ^{*}-1}\left( 1-\nu \right) ^{k}$%
, a distribution displaying an algebraic prefactor (if $\theta ^{*}\neq 1$)
combined to a dominant geometric cutoff. The mean of $K\left( 1\right) $ is $%
\mu ^{*}$ while its variance is $\mu ^{*}/\nu >\mu ^{*}$ (overdispersion
holds). We have 
\begin{equation*}
\frac{\pi _{k+1}^{*}}{\pi _{k}^{*}}=\frac{k+\theta ^{*}}{k+1}\left( 1-\nu
\right) ,
\end{equation*}
so that if $\frac{\pi _{1}^{*}}{\pi _{0}^{*}}>1$ ($\mu ^{*}\nu >1$), the
mode of this distribution is away from zero at about $\left( \mu ^{*}\nu
-1\right) /\nu $; otherwise the mode is at the origin.

The size-biased version of $\pi _{k}^{*}$ is $p_{k}^{*}=k\pi _{k}^{*}/\mu
^{*}$ and the limiting probability $\alpha ^{*}$ that two randomly chosen
individuals from the population are of the same species tends to $0$ like 
\begin{equation*}
\alpha ^{*}:=\sum_{k=1}^{n}\frac{k}{n}p_{k}^{*}=\frac{1}{n\mu ^{*}}%
\sum_{k=1}^{n}k^{2}\pi _{k}^{*}=\frac{1}{p}\left( 1+\frac{1}{\mu ^{*}\nu }%
\right) .
\end{equation*}
Note that under this asymptotic regime, with $Q=Q_{\infty }$, the limiting
number of species present in the population, 
\begin{equation*}
\mathbf{E}\left( Q\right) =\sum_{q=1}^{p}\mathbf{P}\left( K\left( q\right)
>0\right) =p\left( 1-\mathbf{P}\left( K\left( 1\right) =0\right) \right)
\sim p\left( 1-\pi _{0}^{*}\right) =n\frac{1-\nu ^{\theta ^{*}}}{\mu ^{*}}%
\rightarrow \infty .
\end{equation*}
It scales like a fraction of $n$ because $1-\nu ^{\theta ^{*}}<\mu $ as a
result of $-\theta ^{*}\log \nu =\mu ^{*}\nu \log \left( 1/\nu \right)
/\left( 1-\nu \right) <-\log \left( 1-\mu ^{*}\right) $ and $\log \left(
1/\nu \right) <\left( 1-\nu \right) /\nu $ for all $\nu \in \left(
0,1\right) $. Note that as a result, $\pi _{0}^{*}>1-\mu ^{*}$ which is
useful only if $\mu ^{*}\in \left( 0,1\right) $. We will show below that $%
\sigma ^{2}\left( Q\right) \sim p\left( \nu ^{\theta ^{*}}-\nu ^{2\theta
^{*}}\right) $. This asymptotic regime was not considered in \cite{KMG}.%
\newline

\textbf{2.} First fix $n$. If now as in \cite{KMG}, we let $p\rightarrow
\infty $ (infinitely many possible types in the population, see \cite{KC}
for a justification of this in population genetics) and $\beta \rightarrow 0$
(small mutation probability) while $p\beta =\nu >0$ is fixed, then $\mu =%
\frac{n}{p}\rightarrow 0$ and $\theta =\frac{n}{p}\nu /\left( 1-\nu \right)
\rightarrow 0$ while $p\theta \rightarrow n\nu /\left( 1-\nu \right) $. To
the leading order, as $p\rightarrow \infty $%
\begin{equation}
\begin{array}{l}
\pi _{k}\sim \theta \binom{n}{k}\frac{\Gamma \left( k\right) \Gamma \left(
n/\left( 1-\nu \right) -k\right) }{\Gamma \left( n/\left( 1-\nu \right)
\right) }=\pi _{k}^{*}\text{, }k=1,...,n \\ 
\pi _{0}\sim 1-\theta \left( \frac{\Gamma ^{\prime }\left( n/\left( 1-\nu
\right) \right) }{\Gamma \left( n/\left( 1-\nu \right) \right) }-\frac{%
\Gamma ^{\prime }\left( n\nu /\left( 1-\nu \right) \right) }{\Gamma \left(
n\nu /\left( 1-\nu \right) \right) }\right) =\pi _{0}^{*},
\end{array}
\label{f5}
\end{equation}
showing that the equilibrium mass concentrates on state zero: in this low
density regime, the number $n$ of individuals being very few compared to $p$%
, the typical species occupancy is very low.

However (with $\psi \left( z\right) :=\Gamma ^{\prime }\left( z\right)
/\Gamma \left( z\right) $ the digamma function), given $K\left( 1\right)
\geq 1$, for all $k=1,...,n,$%
\begin{equation}
\mathbf{P}\left( K\left( 1\right) =k\mid K\left( 1\right) \geq 1\right) =%
\frac{\pi _{k}^{*}}{1-\pi _{0}^{*}}\sim \binom{n}{k}\frac{B\left( k,n/\left(
1-\nu \right) -k\right) }{\psi \left( n/\left( 1-\nu \right) \right) -\psi
\left( n\nu /\left( 1-\nu \right) \right) }  \label{f6}
\end{equation}
is a well-defined probability mass function as $\theta \rightarrow 0$ ($%
p\rightarrow \infty $) and fixed $n$ and $\nu $: given a species is filled,
it has a well-defined occupancy distribution. Note in passing that this
leads to the non-trivial identity involving the digamma function: for all $%
\nu \in \left( 0,1\right) $%
\begin{equation*}
\psi \left( \frac{n}{1-\nu }\right) -\psi \left( \frac{n\nu }{1-\nu }\right)
=\sum_{k=0}^{n-1}\frac{1}{k+n\nu /\left( 1-\nu \right) }=\sum_{k=1}^{n}%
\binom{n}{k}B\left( k,n/\left( 1-\nu \right) -k\right) .
\end{equation*}
This results from (\ref{f6}) and from $\psi \left( z+1\right) -\psi \left(
z\right) =1/z$ so that by telescopic summation: $\psi \left( z+n\right)
-\psi \left( z\right) =\sum_{k=0}^{n-1}1/\left( k+z\right) .$

With $\left( n\right) _{k}=n!/\left( n-k\right) !$, the mean of $\pi
_{k}^{*} $ is 
\begin{equation*}
\mu =\theta \sum_{k=1}^{n}k\binom{n}{k}\frac{\Gamma \left( k\right) \Gamma
\left( n/\left( 1-\nu \right) -k\right) }{\Gamma \left( n/\left( 1-\nu
\right) \right) }=\theta \frac{1-\nu }{\nu }.
\end{equation*}
It vanishes like $\theta $. However, the size-biased version of $\pi
_{k}^{*} $, namely $p_{k}^{*}=k\pi _{k}^{*}/\mu $, is well-defined, and the
limiting probability $\alpha ^{*}$ that two randomly chosen individuals from
the population are of the same species is (see $4.20$ of \cite{KMG}) 
\begin{equation*}
\alpha ^{*}:=\sum_{k=1}^{n}\frac{k}{n}p_{k}^{*}=\frac{1}{n\mu }%
\sum_{k=1}^{n}k^{2}\pi _{k}^{*}=\frac{1}{1+\nu \left( n-1\right) },
\end{equation*}
which could have been guessed from (\ref{f3a}) as $p\rightarrow \infty $, $%
\nu $ fixed$.$

If now $n$ itself tends to $\infty $ while $\nu $ is still held fixed and $%
n/p\rightarrow 0$ (so that $\theta $ still tends to $0$ as well), owing to $%
\psi \left( z\right) =\frac{\Gamma ^{\prime }\left( z\right) }{\Gamma \left(
z\right) }\underset{z\rightarrow \infty }{\sim }\log z$ and the Stirling
formula, 
\begin{equation}
\mathbf{P}\left( K\left( 1\right) =k\mid K\left( 1\right) \geq 1\right) \sim 
\frac{1}{k}\frac{\left( 1-\nu \right) ^{k}}{\log \left( 1/\nu \right) }\text{%
, }k\geq 1,  \label{f7}
\end{equation}
a Fisher log-series distribution displaying an hyperbolic prefactor combined
to a geometric cutoff, \cite{SR}, \cite{FCW}, \cite{Eng} and \cite{Engen}.
Note again that $\mu =n/p\rightarrow 0$ stipulates that on average each of
the species abundances vanish and only given a species is filled, does it
has a well-defined occupancy distribution.

In this asymptotic regime, with $Q=Q_{\infty }$, the limiting number of
species present in the population, 
\begin{equation}
\mathbf{E}\left( Q\right) \sim p\left( 1-\pi _{0}^{*}\right) \sim p\theta
\log \left( 1/\nu \right) =n\nu \log \left( 1/\nu \right) /\left( 1-\nu
\right) \rightarrow \infty  \label{f7a}
\end{equation}
and it scales like a fraction of $n$ as well (recalling $\log \left( 1/\nu
\right) <\left( 1-\nu \right) /\nu $ for all $\nu \in \left( 0,1\right) $).
From \cite{KMG} 
\begin{equation*}
\sigma ^{2}\left( Q\right) \sim n\left[ \nu \log \left( 1/\nu \right)
/\left( 1-\nu \right) -\nu \right] >0
\end{equation*}
suggesting that $\left( Q-\mathbf{E}\left( Q\right) \right) /\sigma \left(
Q\right) $ is asymptotically normal. Note $\sigma ^{2}\left( Q\right) <%
\mathbf{E}\left( Q\right) $ (underdispersion).\newline

\textbf{3.} Suppose now that $n\rightarrow \infty $, $\nu \rightarrow 0$
while $n\nu =\lambda >0$ is held fixed and $p\nu \rightarrow \infty $. Then $%
\theta =\frac{n}{p}\nu /\left( 1-\nu \right) \sim \frac{\lambda }{p}%
\rightarrow 0$ and, with $k=\left[ nx\right] $%
\begin{equation}
\begin{array}{l}
\pi _{k}\sim \frac{\lambda }{p}\binom{n}{nx}\frac{\Gamma \left( nx\right)
\Gamma \left( n\left( 1-x\right) +\lambda \right) }{\Gamma \left( n+\lambda
\right) }=\frac{\lambda }{pnx}\frac{\Gamma \left( n\left( 1-x\right)
+1+\lambda -1\right) }{\Gamma \left( n\left( 1-x\right) +1\right) }\frac{%
\Gamma \left( n+1\right) }{\Gamma \left( n+1+\lambda -1\right) }\sim \frac{%
\lambda }{pk}\left( 1-\frac{k}{n}\right) ^{\lambda -1}\text{, }k\geq 1 \\ 
\pi _{0}\sim 1-\frac{\lambda }{p}\left( \frac{\Gamma ^{\prime }\left(
n\right) }{\Gamma \left( n\right) }-\frac{\Gamma ^{\prime }\left( \lambda
\right) }{\Gamma \left( \lambda \right) }\right) \sim 1-\frac{\lambda }{p}%
\log n=:\pi _{0}^{*},
\end{array}
\label{f8}
\end{equation}
showing that the equilibrium probability mass concentrates on state zero.
Note that since here $\theta \rightarrow 0$ ($p\rightarrow \infty $) and $%
n\rightarrow \infty $, $\nu \rightarrow 0$ while $n\nu =\lambda >0$, then $%
\mu =n/p\sim \frac{\theta }{\nu }\sim \frac{\lambda }{p\nu }\rightarrow 0$
if $p\nu \rightarrow \infty $ (on average each species abundance vanishes).
With $x\in \left( 0,1\right) $ and $k=\left[ nx\right] \rightarrow \infty $,
putting $n^{-1}=dx$, we have 
\begin{equation*}
\pi _{\left[ nx\right] }^{*}\sim \frac{\lambda }{p}x^{-1}\left( 1-x\right)
^{\lambda -1}dx,
\end{equation*}
not a probability density. Following \cite{KMG} however, $p\pi _{k}^{*}=%
\mathbf{E}\left( N\left( k\right) \right) \sim \frac{\lambda }{k}\left( 1-%
\frac{k}{n}\right) ^{\lambda -1}$ is also the asymptotic expected number of
mutants in the population with $k$ representatives. This shows that in this
regime, the expected number of species whose frequencies range in the
interval $\left( x_{1},x_{2}\right) \subseteq \left[ 0,1\right] $ is $%
\lambda \int_{x_{1}}^{x_{2}}x^{-1}\left( 1-x\right) ^{\lambda -1}dx$ as $%
n\rightarrow \infty .$ Note $\lambda \int_{1/n}^{1}x^{-1}\left( 1-x\right)
^{\lambda -1}dx\sim \lambda \log n\sim p\left( 1-\pi _{0}\right) $ while $%
\lambda \int_{0}^{1}x^{-1}\left( 1-x\right) ^{\lambda -1}dx=\infty .$

With $Q=Q_{\infty }$, the limiting number of species present in the
population, this is consistent with 
\begin{equation}
\mathbf{E}\left( Q\right) =p\left( 1-\pi _{0}\right) \sim \lambda \log
n-\lambda \frac{\Gamma ^{\prime }\left( \lambda \right) }{\Gamma \left(
\lambda \right) }.  \label{ff8}
\end{equation}
In this regime, $\mathbf{E}\left( Q\right) $ scales like $\log n$ and
(mutations being rare) the asymptotic number of types present is sparse
compared to $n$. It is also shown in \cite{KMG} that $\sigma ^{2}\left(
Q\right) \sim \lambda \log n$, so that, upon scaling, $\left( Q-\mathbf{E}%
\left( Q\right) \right) /\sigma \left( Q\right) $ is asymptotically normal.%
\newline

\textbf{4.} The authors of \cite{KMG} also consider the asymptotic regime
for which $n\rightarrow \infty $, $\nu \rightarrow 0$ while $\nu n\log
n=c\geq 0$ ($\lambda =c/\log n\rightarrow 0$) for which from the above
estimates and $\lambda \frac{\Gamma ^{\prime }\left( \lambda \right) }{%
\Gamma \left( \lambda \right) }\underset{\lambda \rightarrow 0^{+}}{\sim }%
-1, $ $\mathbf{E}\left( Q\right) \sim 1+c$ and $\sigma ^{2}\left( Q\right)
\sim c.$ In this asymptotic regime, only a finite number of types are
present.\newline

\textbf{5.} (the dense case). If $p$ is fixed and $n\rightarrow \infty $, $%
\nu \rightarrow 0$ while $n\nu =\lambda >0$, then $\theta =\frac{n}{p}\nu
/\left( 1-\nu \right) \rightarrow \theta ^{*}=\lambda /p>0$ and 
\begin{equation}
\begin{array}{l}
\pi _{k}\sim \frac{\Gamma \left( \lambda \right) }{\Gamma \left( \theta
^{*}\right) \Gamma \left( \lambda -\theta ^{*}\right) }\frac{\Gamma \left(
k+1+\theta ^{*}-1\right) }{\Gamma \left( k+1\right) }\frac{\Gamma \left(
n-k+1+\lambda -\theta ^{*}-1\right) }{\Gamma \left( n-k+1\right) }\frac{%
\Gamma \left( n+1\right) }{\Gamma \left( n+1+\lambda -1\right) } \\ 
\sim \frac{\Gamma \left( \lambda \right) }{\Gamma \left( \theta ^{*}\right)
\Gamma \left( \lambda -\theta ^{*}\right) }\frac{\Gamma \left( k+1+\theta
^{*}-1\right) }{\Gamma \left( k+1\right) }\frac{\left( n-k+1\right)
^{\lambda -\theta ^{*}-1}}{\left( n+1\right) ^{\lambda -1}}=\pi _{k}^{*}.
\end{array}
\label{ff9}
\end{equation}
If $k=\left[ nx\right] \rightarrow \infty $ with $x\in \left( 0,1\right) $%
\begin{equation}
\begin{array}{l}
\pi _{\left[ nx\right] }^{*}\sim n^{-1}n\frac{\Gamma \left( \lambda \right) 
}{\Gamma \left( \theta ^{*}\right) \Gamma \left( \lambda -\theta ^{*}\right) 
}\left( nx\right) ^{\theta ^{*}-1}\left( n\left( 1-x\right) \right)
^{\lambda -\theta ^{*}-1}n^{-\left( \lambda -1\right) } \\ 
\sim dx\frac{\Gamma \left( \lambda \right) }{\Gamma \left( \theta
^{*}\right) \Gamma \left( \lambda -\theta ^{*}\right) }x^{\theta
^{*}-1}\left( 1-x\right) ^{\lambda -\theta ^{*}-1},
\end{array}
\label{ff10}
\end{equation}
a beta density with parameters $\theta ^{*}=\lambda /p,\lambda -\theta
^{*}=\lambda \left( 1-1/p\right) .$ This shows (with $n^{-1}=dx$) that, in
this asymptotic regime, $n^{-1}K\left( 1\right) \overset{d}{\rightarrow }$ $%
B\left( \theta ^{*},\lambda -\theta ^{*}\right) $ as $n\rightarrow \infty .$
This asymptotic regime was not considered in \cite{KMG} either.

Note that, from (\ref{f3a}), the limiting probability $\alpha ^{*}$ that two
randomly chosen individuals from the population are of the same species is 
\begin{equation*}
\alpha =\frac{p+\nu \left( n-1\right) }{p\left( 1+\nu \left( n-1\right)
\right) }\rightarrow \alpha ^{*}=\frac{p+\lambda }{p\left( 1+\lambda \right) 
},
\end{equation*}
approaching $1/\left( 1+\lambda \right) $ if $p$ is in turn large enough.%
\newline

\textbf{Remark:} Regime \textbf{1} deals with a large population of size $n$
together with a large number of types $p$ both of the same order of
magnitude (a case with asymptotic density $n/p\rightarrow \mu ^{*}$). It is
balanced. In the regimes \textbf{2} to \textbf{4}, $n\ll p$, a dilute phase
situation with low density of individuals compared to the species range. And
while scrolling from regimes \textbf{2} to \textbf{4, }$n\nu $ ranges from
infinity to zero, through moderate in regime \textbf{3}. The main results
are from \cite{KMG}. In the dense (large density) regime \textbf{5} on the
contrary, $n\gg p$ and the population is made of few types but a large
number of individuals. It is sometimes adapted to the surname distribution
studies: for instance in France, there are about $p=1.5$ million different
surnames for a population of about $n=67$ millions people. As of $2000$,
about $p=286$ Korean family names were reported in use in South Korea for a
population around $n=50$ millions people. In both cases however, $n$ cannot
be assumed having stabilized. The study \cite{Zei} dealing with the
Sardinian island looks convincing. Note that the whole KMG theory breaks
down would the hypothesis of a constant population size be relaxed, as in
the Yule approach to the speciation process briefly addressed in the
introduction. A hint of the drastic changes to be made in the neutral
context when the population size is held constant on average only is to be
found in \cite{Hui3}. Note also that there is no ``selection effect'' in the
model, the adjunction of which would also considerably alter the KMG
machinery (\cite{KMG} p. $422$).

\section{Joint distributions of species abundances under KMG mutation model}

So far we only obtained useful information on the limiting occupancy of a
typical species $K\left( 1\right) $ and only partial (mean and variance)
information on the asymptotic number $Q$ of filled species. We are able to
be more precise. We start with fixed $n$ and $p$ before considering
asymptotic regimes.

With $\theta =\frac{n}{p}\nu /\left( 1-\nu \right) $, consider the Dirichlet
continuous density function, say $D_{p}\left( \theta \right) $, on the
simplex $\left\{ s_{q}\in \left( 0,1\right) :\sum_{q=1}^{p}s_{q}=1\right\} $%
\begin{equation}
f_{S_{1},...,S_{p}}\left( s_{1},...,s_{p}\right) =\frac{\Gamma \left(
p\theta \right) }{\Gamma \left( \theta \right) ^{p}}\prod_{q=1}^{p}s_{q}^{%
\theta -1}\cdot \delta _{\left( \sum_{q=1}^{p}s_{q}-1\right) }.  \label{eq1}
\end{equation}
The law of $\mathbf{S}_{p}:=\left( S_{1},...,S_{p}\right) $ can as well be
characterized by its joint moment function ($\lambda _{q}>0$) 
\begin{equation}
\mathbf{E}\left( \prod_{q=1}^{p}S_{q}^{\lambda _{q}}\right) =\frac{1}{\left[
p\theta \right] _{\sum_{q=1}^{p}\lambda _{q}}}\prod_{q=1}^{p}\left[ \theta
\right] _{\lambda _{q}}.  \label{eq1a}
\end{equation}
where $\left[ \theta \right] _{\lambda }=\Gamma \left( \theta +\lambda
\right) /\Gamma \left( \theta \right) $.

Clearly, the equilibrium joint distribution of $\mathbf{K}_{t}$, namely $%
\mathbf{K}:=\left( K\left( q\right) ,q=1,...,p\right) $, is a $D_{p}\left(
\theta \right) $ $\mathbf{s}$-mixture of a multinomial multin$\left( n,%
\mathbf{s}\right) $ distribution where $\mathbf{s=}\left(
s_{1},...,s_{p}\right) $. With $\Bbb{N}_{0}:=\left\{ 0,1,2,...\right\} $, it
is thus a Dirichlet-multinomial distribution on the now discrete simplex $%
\left\{ k_{q}\in \Bbb{N}_{0}:\sum_{q=1}^{p}k_{q}=n\right\} $ with (see \cite
{Hui2}, Theorem $6$, for instance) 
\begin{equation}
\mathbf{P}\left( \mathbf{K}=\mathbf{k}\right) =\mathbf{EP}\left( \mathbf{K}=%
\mathbf{k}\mid \mathbf{S}_{p}\right) =\frac{n!}{\left[ p\theta \right] _{n}}%
\prod_{q=1}^{p}\frac{\left[ \theta \right] _{k_{q}}}{k_{q}!}.  \label{f8a}
\end{equation}
Here $\mathbf{K\mid S}_{p}\overset{d}{\sim }$ multin$\left( n,\mathbf{S}%
_{p}\right) $ and $\mathbf{S}_{p}\overset{d}{\sim }$ $D_{p}\left( \theta
\right) .$ It is an exchangeable distribution, each margin being identically
distributed, but of course, owing to $\sum_{q=1}^{p}K\left( q\right) =n$,
the $K\left( q\right) $'s are not independent. We observe that,
equivalently, with all $u_{q}\in \left( 0,1\right) $, the joint probability
generating function of $\mathbf{K}$ is 
\begin{equation}
\mathbf{E}\left( \prod_{q=1}^{p}u_{q}^{K\left( q\right) }\right) =\mathbf{E}%
\left[ \left( \sum_{q=1}^{p}u_{q}S_{q}\right) ^{n}\right]  \label{ff11}
\end{equation}
from which joint statistical information can be extracted using moment
identities of the Dirichlet distribution. The simplest one is the (negative)
covariance between any two pairs $\left( K\left( 1\right) ,K\left( 2\right)
\right) $ of equilibrium species abundances which can easily be found to be
from (\ref{ff11}) and using (\ref{eq1a}) 
\begin{equation*}
\text{Cov}\left( K\left( 1\right) ,K\left( 2\right) \right) =-\frac{\sigma
^{2}\left( K\left( 1\right) \right) }{p-1}=-\frac{n}{p}\left( \frac{n}{p}-%
\frac{\left( n-1\right) \theta }{p\theta +1}\right) =-\left( \frac{n}{p}%
\right) ^{2}\frac{1}{1+\nu \left( n-1\right) }.
\end{equation*}

Coming back to (\ref{f8a}), it is convenient to introduce the related joint
probability 
\begin{equation}
\mathbf{P}\left( K\left( 1\right) =k_{1},...,K\left( q\right) =k_{q}\mathbf{;%
}Q=q\right) =\binom{p}{q}\frac{n!}{\left[ p\theta \right] _{n}}%
\prod_{q^{\prime }=1}^{q}\frac{\left[ \theta \right] _{k_{q^{\prime }}}}{%
k_{q^{\prime }}!}  \label{f9}
\end{equation}
where $1\leq q\leq p$ and then with $\sum_{q^{\prime }=1}^{q}k_{q^{\prime
}}=n$ and all $k_{q^{\prime }}\geq 1.$ It is the joint probability that
there are $q\in \left\{ 1,...,p\right\} $ \textbf{filled} species cells and
that $\left( k_{1},...,k_{q}\right) $ are their effective abundance
occupancies. Letting $\sigma _{n}\left( \theta \right) :=n!\left[
x^{n}\right] Z_{\theta }\left( x\right) =\left[ \theta \right] _{n}$ where $%
Z_{\theta }\left( x\right) =e^{\theta \phi \left( x\right) }$ and $\phi
\left( x\right) =-\log \left( 1-x\right) $, with $\Bbb{N}:=\left\{
1,2,...\right\} $, $\mathbf{k}_{q}:=\left( k_{1},...,k_{q}\right) $ and $%
\left| \mathbf{k}_{q}\right| =\sum_{q^{\prime }=1}^{q}k_{q^{\prime }}$, we
have\emph{\ } 
\begin{equation}
\mathbf{P}\left( Q=q\right) =\binom{p}{q}\frac{n!}{\sigma _{n}\left( p\theta
\right) }\sum_{\mathbf{k}_{q}\in \Bbb{N}^{q}:\text{ }\left| \mathbf{k}%
_{q}\right| =n}\prod_{q^{\prime }=1}^{q}\frac{\sigma _{k_{q^{\prime
}}}\left( \theta \right) }{k_{q^{\prime }}!}\text{, }q=1,...,p.  \label{f9c}
\end{equation}
The expression (\ref{f9}) turns out to be the canonical Gibbs distribution
on the simplex $\left\{ \mathbf{k}_{q}\in \Bbb{N}^{q}:\text{ }\left| \mathbf{%
k}_{q}\right| =n\right\} $, the finite size-$p$\ partitions of $n$\ into $q$%
\ distinct clusters (the filled species). In this language, the normalizing
quantity $\sigma _{n}\left( p\theta \right) /n!$\ is called the canonical
Gibbs partition function.\emph{\newline
}

Now, from (\ref{f9c}), with $\left( p\right) _{q}:=p!/\left( p-q\right) !$%
\begin{equation}
\mathbf{P}\left( Q=q\right) =\frac{\left( p\right) _{q}}{\sigma _{n}\left(
p\theta \right) }B_{n,q}\left( \sigma _{\bullet }\left( \theta \right)
\right) ,\text{ }q\in \left\{ 1,...,p\wedge n\right\} ,  \label{f9a}
\end{equation}
where 
\begin{equation}
B_{n,q}\left( \sigma _{\bullet }\left( \theta \right) \right) :=\frac{n!}{q!}%
\sum_{\mathbf{k}_{q}\in \Bbb{N}^{q}:\text{ }\left| \mathbf{k}_{q}\right| =n%
\text{ }}\prod_{q^{\prime }=1}^{q}\frac{\sigma _{k_{q^{\prime }}}\left(
\theta \right) }{k_{q^{\prime }}!}=\frac{n!}{q!}\left[ x^{n}\right] \left(
Z_{\theta }\left( x\right) -1\right) ^{q}  \label{f10}
\end{equation}
are the Bell polynomials in the polynomial variables $\sigma _{\bullet
}\left( \theta \right) :=\left( \sigma _{1}\left( \theta \right) ,\sigma
_{2}\left( \theta \right) ,...\right) $, \cite{Comtet}$.$ Here again $%
Z_{\theta }\left( x\right) =\left( 1-x\right) ^{-\theta }$ and $\sigma
_{n}\left( \theta \right) =n!\left[ z^{n}\right] e^{-\theta \log \left(
1-z\right) }=\left[ \theta \right] _{n}.$

Conditioning the canonical Gibbs distribution on the number of filled
species being equal to $q$ yields the corresponding micro-canonical
distribution as 
\begin{equation}
\mathbf{P}\left( K\left( 1\right) =k_{1},...,K\left( q\right) =k_{q}\mid
Q=q\right) =\frac{n!}{q!}\frac{1}{B_{n,q}\left( \sigma _{\bullet }\left(
\theta \right) \right) }\prod_{q^{\prime }=1}^{q}\frac{\sigma _{k_{q^{\prime
}}}\left( \theta \right) }{k_{q^{\prime }}!}.  \label{fg9}
\end{equation}
The new normalizing constant $B_{n,q}\left( \sigma _{\bullet }\left( \theta
\right) \right) /n!$ may be called the microcanonical partition function.
The special feature of the occupancy distributions (\ref{f9}), (\ref{f9a})
and (\ref{fg9}) is that $\theta =\frac{n}{p}\nu /\left( 1-\nu \right) $
depends on $n$, $p$ and $\nu .$\newline

Let us now first characterize the full distribution of $Q$ which depends on $%
n$, $p$ and $\nu $ (via $\theta $) before any asymptotics is considered. We
have:

$\left( a\right) $\emph{\ Assume }$n\geq p$\emph{. With }$u\in \left[
0,1\right] $\emph{, the probability generating function of }$Q$\emph{\ is
given by } 
\begin{equation}
\mathbf{E}\left( u^{Q}\right) =\sum_{q=0}^{p-1}\binom{p}{q}u^{p-q}\left(
1-u\right) ^{q}\frac{\sigma _{n}\left( \left( p-q\right) \theta \right) }{%
\sigma _{n}\left( p\theta \right) },\text{ with}  \label{f10a}
\end{equation}
\begin{equation}
\mathbf{P}\left( Q=q\right) =\frac{\binom{p}{q}}{\sigma _{n}\left( p\theta
\right) }\sum_{q^{\prime }=1}^{q}\left( -1\right) ^{q-q^{\prime }}\binom{q}{%
q^{\prime }}\sigma _{n}\left( q^{\prime }\theta \right) ,\text{ }q\in
\left\{ 1,...,p\right\} .  \label{f10b}
\end{equation}
\emph{In addition, } 
\begin{equation}
\mathbf{E}\left( Q\right) =p\left( 1-\frac{\sigma _{n}\left( \left(
p-1\right) \theta \right) }{\sigma _{n}\left( p\theta \right) }\right) \text{
and}  \label{f10b2}
\end{equation}
\begin{equation}
\sigma ^{2}\left( Q\right) =p\left( \frac{\sigma _{n}\left( \left(
p-1\right) \theta \right) }{\sigma _{n}\left( p\theta \right) }+\left(
p-1\right) \frac{\sigma _{n}\left( \left( p-2\right) \theta \right) }{\sigma
_{n}\left( p\theta \right) }-p\left( \frac{\sigma _{n}\left( \left(
p-1\right) \theta \right) }{\sigma _{n}\left( p\theta \right) }\right)
^{2}\right) .  \label{f10b3}
\end{equation}

$\left( b\right) $\emph{\ If }$n<p$\emph{, (\ref{f10a}) and (\ref{f10b})
still hold, but now with a modified support for }$Q$\emph{'s law:\ } 
\begin{equation}
\mathbf{P}\left( Q=q\right) =\frac{\binom{p}{q}}{\sigma _{n}\left( p\theta
\right) }\sum_{q^{\prime }=1}^{q}\left( -1\right) ^{q-q^{\prime }}\binom{q}{%
q^{\prime }}\sigma _{n}\left( q^{\prime }\theta \right) ,\text{ }q\in
\left\{ 1,...,n\right\} .  \label{f10b1}
\end{equation}

Statement $\left( a\right) $ follows from $B_{n,q}\left( \sigma _{\bullet
}\left( \theta \right) \right) =\frac{n!}{q!}\left[ x^{n}\right] \left(
Z_{\theta }\left( x\right) -1\right) ^{q}.$ Indeed, from (\ref{f9a}) 
\begin{equation*}
\begin{array}{l}
\mathbf{E}\left( u^{Q}\right) =\sum_{q=0}^{p}u^{q}\left( p\right) _{q}\frac{%
B_{n,q}\left( \sigma _{\bullet }\left( \theta \right) \right) }{\sigma
_{n}\left( p\theta \right) }=\frac{n!}{\sigma _{n}\left( p\theta \right) }%
\sum_{q=0}^{p}\binom{p}{q}\left[ x^{n}\right] \left( u\left( Z_{\theta
}\left( x\right) -1\right) \right) ^{q} \\ 
=\frac{n!}{\sigma _{n}\left( p\theta \right) }\left[ x^{n}\right] \left(
1-u+uZ_{\theta }\left( x\right) \right) ^{p}=\frac{n!}{\sigma _{n}\left(
p\theta \right) }\sum_{q=0}^{p}\binom{p}{q}u^{p-q}\left( 1-u\right)
^{q}\left[ x^{n}\right] Z_{\theta }\left( x\right) ^{p-q} \\ 
=\sum_{q=0}^{p-1}\binom{p}{q}u^{p-q}\left( 1-u\right) ^{q}\frac{\sigma
_{n}\left( \left( p-q\right) \theta \right) }{\sigma _{n}\left( p\theta
\right) }.
\end{array}
\end{equation*}
The alternating sum expression of $\mathbf{P}\left( Q=q\right) $ follows
from extracting $\left[ u^{q}\right] \mathbf{E}\left( u^{Q}\right) $ and the
mean and variance of $Q$ from the evaluations of the first and second
derivatives of $\mathbf{E}\left( u^{Q}\right) $ with respect to $u$ at $u=1.$

Statement $\left( b\right) $ follows from similar considerations. Indeed, in
principle, we should start with $\mathbf{E}\left( u^{Q}\right)
=\sum_{q=0}^{n}u^{q}\left( p\right) _{q}\frac{B_{n,q}\left( \sigma _{\bullet
}\left( \theta \right) \right) }{\sigma _{n}\left( p\theta \right) }$ where
the $q-$sum now stops at $q=n=p\wedge n$. But the upper bound of this $q-$%
sum can be extended to $p$ because $B_{n,q}\left( \sigma _{\bullet }\left(
\theta \right) \right) =0$ if $q>n$. \newline

In the particular mutation case discussed here, $\sigma _{n}\left( \theta
\right) =\left[ \theta \right] _{n}=\theta \left( \theta +1\right) ...\left(
\theta +n-1\right) =\Gamma \left( \theta +n\right) /\Gamma \left( \theta
\right) $ (the Ewens-Dirichlet model, \cite{Ewens}). From (\ref{f10b}), (\ref
{f10b2}), for instance, 
\begin{equation*}
\mathbf{P}\left( Q=1\right) =p\frac{\sigma _{n}\left( \theta \right) }{%
\sigma _{n}\left( p\theta \right) }=p\frac{\left[ \theta \right] _{n}}{%
\left[ p\theta \right] _{n}}=p\frac{\Gamma \left( \theta +n\right) }{\Gamma
\left( \theta \right) }\frac{\Gamma \left( p\theta \right) }{\Gamma \left(
p\theta +n\right) }.
\end{equation*}

\begin{equation*}
\mathbf{E}\left( Q\right) =p\left( 1-\frac{\left[ \left( p-1\right) \theta
\right] _{n}}{\left[ p\theta \right] _{n}}\right) =p\left( 1-\frac{\Gamma
\left( \left( p-1\right) \theta +n\right) }{\Gamma \left( \left( p-1\right)
\theta \right) }\frac{\Gamma \left( p\theta \right) }{\Gamma \left( p\theta
+n\right) }\right)
\end{equation*}
We now illustrate some of the consequences of the latter expressions under
three asymptotic regimes discussed earlier.\newline

Regime \textbf{1.} If $n$ and $p\rightarrow \infty $ while $n/p\rightarrow
\mu ^{*}$ as in the balanced regime \textbf{1}$,$ then $\theta =\frac{n}{p}%
\nu /\left( 1-\nu \right) \rightarrow \theta ^{*}=\mu ^{*}\nu /\left( 1-\nu
\right) $ and, from (\ref{f10b2}), in a consistent way with previous
results, 
\begin{equation*}
\begin{array}{l}
\mathbf{E}\left( Q\right) \sim p\left( 1-\frac{\Gamma \left( p\left( \theta
^{*}+\mu ^{*}\right) -\theta ^{*}\right) }{\Gamma \left( p\left( \theta
^{*}+\mu ^{*}\right) \right) }\frac{\Gamma \left( p\theta ^{*}\right) }{%
\Gamma \left( p\theta ^{*}-\theta ^{*}\right) }\right) \\ 
\sim p\left( 1-\frac{\left( p\left( \theta ^{*}+\mu ^{*}\right) \right)
^{-\theta ^{*}}}{\left( p\theta ^{*}\right) ^{-\theta ^{*}}}\right) =p\left(
1-\left( \frac{\theta ^{*}}{\theta ^{*}+\mu ^{*}}\right) ^{\theta
^{*}}\right) =p\left( 1-\nu ^{\theta ^{*}}\right) .
\end{array}
\end{equation*}
Proceeding similarly, from (\ref{f10b3}) $\sigma ^{2}\left( Q\right) \sim
p\left( \nu ^{\theta ^{*}}-\nu ^{2\theta ^{*}}\right) $, suggesting $\left(
Q-\mathbf{E}\left( Q\right) \right) /\sigma \left( Q\right) $ is
asymptotically normal in regime \textbf{1 }as well.

From (\ref{ff11}) and (\ref{eq1a}), with $1\leq q<p$, $a=\nu /\left( 1-\nu
\right) $ and $p\theta \sim na$ and $\theta \sim \theta ^{*}=\mu ^{*}a$%
\begin{equation*}
\begin{array}{l}
\mathbf{E}\left( \prod_{q^{\prime }=1}^{q}u_{q}^{K\left( q^{\prime }\right)
}\right) =\mathbf{E}\left[ \left( 1+\sum_{q^{\prime }=1}^{q}\left(
u_{q^{\prime }}-1\right) S_{q^{\prime }}\right) ^{n}\right] \\ 
=\sum_{k_{1}+...+k_{q+1}=n}\binom{n}{k_{1}...k_{q+1}}\prod_{q^{\prime
}=1}^{q}\left( u_{q^{\prime }}-1\right) ^{k_{q^{\prime }}}\mathbf{E}\left(
\prod_{q^{\prime }=1}^{q}S_{q^{\prime }}^{k_{q^{\prime }}}\right) \\ 
=\sum_{k_{1}+...+k_{q+1}=n}\binom{n}{k_{1}...k_{q+1}}\prod_{q^{\prime
}=1}^{q}\left( u_{q^{\prime }}-1\right) ^{k_{q^{\prime }}}\frac{%
\prod_{q^{\prime }=1}^{q}\left[ \theta \right] _{k_{q^{\prime }}}}{\left[
p\theta \right] _{\sum_{q^{\prime }=1}^{q}k_{q^{\prime }}}} \\ 
=\sum_{k_{q+1}^{\prime }=0}^{n}\frac{1}{\left( n-k_{q+1}^{\prime }\right) !}%
\sum_{k_{1}+...+k_{q}=k_{q+1}^{\prime }}\binom{n}{k_{1}...k_{q}}%
\prod_{q^{\prime }=1}^{q}\left( u_{q^{\prime }}-1\right) ^{k_{q^{\prime }}}%
\frac{\prod_{q^{\prime }=1}^{q}\left[ \theta \right] _{k_{q^{\prime }}}}{%
\left[ p\theta \right] _{k_{q+1}^{\prime }}} \\ 
\sim \sum_{k_{q+1}^{\prime }=0}^{n}\binom{n}{k_{q+1}^{\prime }}\left(
na\right) ^{-k_{q+1}^{\prime }}\sum_{k_{1}+...+k_{q}=k_{q+1}^{\prime }}%
\binom{k_{q+1}^{\prime }}{k_{1}...k_{q}}\prod_{q^{\prime }=1}^{q}\left(
\left[ \theta ^{*}\right] _{k_{q^{\prime }}}\left( u_{q^{\prime }}-1\right)
^{k_{q^{\prime }}}\right) \\ 
\sim \sum_{k_{q+1}^{\prime }=0}^{\infty }a^{-k_{q+1}^{\prime
}}\sum_{k_{1}+...+k_{q}=k_{q+1}^{\prime }}\prod_{q^{\prime }=1}^{q}\left(
\left[ \theta ^{*}\right] _{k_{q^{\prime }}}\left( u_{q^{\prime }}-1\right)
^{k_{q^{\prime }}}\right) /k_{q^{\prime }}!,
\end{array}
\end{equation*}
the pgf of the multivariate negative binomial distribution of $\left(
K\left( q^{\prime }\right) ;q^{\prime }=1,...,q\right) $. Note that if $q=1$%
, this pgf reduces, as required from Section $3$, to 
\begin{equation*}
\mathbf{E}\left( u_{1}^{K\left( 1\right) }\right) =\sum_{k=0}^{\infty }a^{-k}%
\frac{\left[ \theta ^{*}\right] _{k}}{k!}\left( u_{1}-1\right) ^{k}=\left( 
\frac{\nu }{1-\left( 1-\nu \right) u_{1}}\right) ^{\theta ^{*}}.
\end{equation*}
\newline

Regime \textbf{2.} (infinitely many possible types in the population). First
fix population size $n$. If now as in regime \textbf{2}, we let $%
p\rightarrow \infty $ and $\beta \rightarrow 0$ (small mutation probability)
while $p\beta =\nu >0$ is fixed, then $\theta =\frac{n}{p}\nu /\left( 1-\nu
\right) \rightarrow 0$ while $p\theta \sim n\nu /\left( 1-\nu \right)
=:\gamma $.

Recall $\sigma _{n}\left( \theta \right) :=n!\left[ x^{n}\right] Z_{\theta
}\left( x\right) =\left[ \theta \right] _{n}$ where $Z_{\theta }\left(
x\right) =e^{\theta \phi \left( x\right) },$ $\phi \left( x\right) =-\log
\left( 1-x\right) $ and $\phi _{i}=\left[ x^{i}\right] \phi \left( x\right)
=\left( i-1\right) !.$ We have $B_{n,q}\left( \phi _{\bullet }\right) =\frac{%
n!}{q!}\left[ x^{n}\right] \phi \left( x\right) ^{q}=s_{n,q}$, the first
kind absolute Stirling numbers, \cite{Comtet}$.$

When $\theta \rightarrow 0$, , from (\ref{f10}), $B_{n,q}\left( \sigma
_{\bullet }\left( \theta \right) \right) =\frac{n!}{q!}\left[ x^{n}\right]
\left( Z_{\theta }\left( x\right) -1\right) ^{q}\sim \frac{n!}{q!}\theta
^{q}\left[ x^{n}\right] \phi \left( x\right) ^{q}=\theta ^{q}B_{n,q}\left(
\phi _{\bullet }\right) =\theta ^{q}s_{n,q}$. Thus, recalling $p\theta \sim
\gamma ,$ (\ref{f9a}) becomes 
\begin{equation}
\mathbf{P}\left( Q=q\right) =\frac{\left( p\right) _{q}}{\sigma _{n}\left(
p\theta \right) }B_{n,q}\left( \sigma _{\bullet }\left( \theta \right)
\right) \sim \frac{\left( p\theta \right) ^{q}s_{n,q}}{\sigma _{n}\left(
p\theta \right) }=\frac{\gamma ^{q}s_{n,q}}{\sigma _{n}\left( \gamma \right) 
}\text{, }q=1,...,n\text{,}  \label{equ1}
\end{equation}
giving the simple asymptotic shape of the law of $Q$ for a size $n$
population with infinitely many types. It depends on $\nu $, via $\gamma
=n\nu /\left( 1-\nu \right) $. The probability generating function of this
limiting $Q$ is 
\begin{equation*}
\mathbf{E}\left( u^{Q}\right) =\frac{\sigma _{n}\left( \gamma u\right) }{%
\sigma _{n}\left( \gamma \right) }=\frac{\left[ \gamma u\right] _{n}}{\left[
\gamma \right] _{n}}=u\prod_{q=1}^{n-2}\left( \frac{\gamma u+q}{\gamma +q}%
\right) ,
\end{equation*}
showing that $Q\overset{d}{=}1+\sum_{q=1}^{n-2}B_{q}$ where the $B_{q}$'s
are independent Bernoulli random variables with success parameters $\frac{%
\gamma }{\gamma +q}$ where $\gamma =n\nu /\left( 1-\nu \right) $. Recalling $%
\psi \left( z\right) \underset{z\rightarrow \infty }{\sim }\log z$ and
because 
\begin{equation*}
\sum_{q=0}^{n-1}\frac{\gamma }{\gamma +q}=\frac{n\nu }{1-\nu }\left( \psi
\left( \frac{n}{1-\nu }\right) -\psi \left( \frac{n\nu }{1-\nu }\right)
\right) ,
\end{equation*}
$\sum_{q=0}^{n-1}\frac{\gamma }{\gamma +q}\sim \frac{n\nu }{1-\nu }\log
\left( 1/\nu \right) $ and by strong law of large numbers $Q/n\underset{%
n\rightarrow \infty }{\rightarrow }\nu \log \left( 1/\nu \right) /\left(
1-\nu \right) $ almost surely (completing (\ref{f7a})).\footnote{%
This formalism resembles the one of the asymptotic number $Q$ of filled
tables in the Chinese restaurant problem with $n$ customers (see Sections $%
3.1$ and $3.2$ of \cite{Pit}). However its asymptotic behavior is of a
different nature because here $\gamma $ depends on $n$ (and $\nu $), leading
to $Q$ of order $n$ rather than $\log n$.}

Owing now to $\sigma _{k_{q^{\prime }}}\left( \theta \right) =\left[ \theta
\right] _{k_{q^{\prime }}}\sim \theta \Gamma \left( k_{q^{\prime }}\right)
=\theta \left( k_{q^{\prime }}-1\right) !$, with the $k_{q}$'s positive
summing to $n,$ from (\ref{f9}), 
\begin{equation*}
\begin{array}{l}
\mathbf{P}\left( K\left( 1\right) =k_{1},...,K\left( q\right) =k_{q}\mathbf{;%
}Q=q\right) =\binom{p}{q}\frac{n!}{\left[ p\theta \right] _{n}}%
\prod_{q^{\prime }=1}^{q}\frac{\left[ \theta \right] _{k_{q^{\prime }}}}{%
k_{q^{\prime }}!} \\ 
\sim \frac{n!}{q!}\frac{\gamma ^{q}}{\left[ \gamma \right] _{n}}%
\prod_{q^{\prime }=1}^{q}\frac{1}{k_{q^{\prime }}!},
\end{array}
\end{equation*}
and from (\ref{fg9}) 
\begin{equation*}
\begin{array}{l}
\mathbf{P}\left( K\left( 1\right) =k_{1},...,K\left( q\right) =k_{q}\mid
Q=q\right) =\frac{n!}{q!}\frac{1}{B_{n,q}\left( \sigma _{\bullet }\left(
\theta \right) \right) }\prod_{q^{\prime }=1}^{q}\frac{\sigma _{k_{q^{\prime
}}}\left( \theta \right) }{k_{q^{\prime }}!} \\ 
\sim \frac{n!}{q!}\frac{1}{\theta ^{q}s_{n,q}}\prod_{q^{\prime }=1}^{q}\frac{%
\theta \left( k_{q^{\prime }}-1\right) !}{k_{q^{\prime }}!}=\frac{n!}{q!}%
\frac{1}{s_{n,q}}\prod_{q^{\prime }=1}^{q}\frac{1}{k_{q^{\prime }}},
\end{array}
\end{equation*}
the Ewens sampling formula, \cite{Ewens}, \cite{TE}. This gives the
asymptotic shape of the joint equilibrium species abundance vector, given $q$
of them are represented at equilibrium. A curious feature of this last
distribution is that it is independent of $\nu $.\newline

Regime \textbf{3}. If as here $n\rightarrow \infty $ and $\nu \rightarrow 0$
while $n\nu =\lambda $, then $\gamma \sim \gamma ^{*}=\lambda $ and we are
now in the asymptotic region akin to the Chinese restaurant process. For
instance 
\begin{equation}
\sum_{q=0}^{n-1}\frac{\gamma }{\gamma +q}\overset{*}{\sim }\sum_{q=0}^{n-1}%
\frac{\lambda }{\lambda +q}\sim \lambda \log n  \label{Equ2}
\end{equation}
and $Q/\log n\underset{n\rightarrow \infty }{\rightarrow }\lambda $ almost
surely (completing (\ref{ff8})). The analogy is thus with a `chinese'
restaurant with infinitely many indistinguishable tables, each of which has
infinite capacity. In the table filling process, the first customer sits at
some table while the next one either sits at the same table or at a
different one. The process continues, with each customer choosing either to
sit at an occupied table with a probability proportional to the number of
customers already there or at some already unoccupied table. Under the
condition of Regime $3$, after step $n$, the occupancies of the tables are
given by (\ref{f9}) and the $n$ customers will be partitioned among $Q\leq n$
filled tables (or blocks of the partition) with $Q$ of order $\log n$. The
outcomes of this process are exchangeable as the order in which the
customers sit does not affect the probability of the final distribution.

If, as in regime \textbf{4}, $n\rightarrow \infty $ and $\nu \rightarrow 0$
while $n\nu \sim c/\log n$ for some $c>0$, then $\gamma \sim \gamma
^{*}=c/\log n$ and $\mathbf{E}\left( Q\right) =\sum_{q=0}^{n-1}\frac{\gamma 
}{\gamma +q}\overset{*}{\sim }1+c.$ It is easy to check that here $Q-1%
\overset{*}{\rightarrow }$Poi$\left( c\right) ,$ a Poisson random variable
with mean $c$.\newline

Regime \textbf{5.} (Finitely many types and very large population size).
Finally, from (\ref{ff11}), with $X_{q}$, $q=1,...,p$ iid Gamma$\left(
\theta \right) $ random variables and $\widetilde{X}_{q}:=X_{q}/%
\sum_{q=1}^{p}X_{q}$ and exploiting the Gamma structure of Dirichlet
distributions, 
\begin{equation*}
\begin{array}{l}
\mathbf{E}\left( \prod_{q=1}^{p}u_{q}^{K\left( q\right) /n}\right) =\mathbf{E%
}\left[ \left( \sum_{q=1}^{p}u_{q}^{1/n}S_{q}\right) ^{n}\right] =\frac{1}{%
\left[ p\theta \right] _{n}}\mathbf{E}\left[ \left(
\sum_{q=1}^{p}u_{q}^{1/n}X_{q}\right) ^{n}\right] \\ 
\underset{n\uparrow \infty }{\sim }\frac{1}{\left[ p\theta \right] _{n}}%
\mathbf{E}\left[ \left( \sum_{q=1}^{p}X_{q}\right) ^{n}\left( 1+\frac{1}{n}%
\sum_{q=1}^{p}\widetilde{X}_{q}\log u_{q}\right) ^{n}\right] \\ 
\underset{n\uparrow \infty }{\sim }\mathbf{E}\left( \prod_{q=1}^{p}u_{q}^{%
\widetilde{X}_{q}}\right) =\mathbf{E}\left(
\prod_{q=1}^{p}u_{q}^{S_{q}}\right) .
\end{array}
\end{equation*}
Thus, generalizing (\ref{ff10}), 
\begin{equation}
\mathbf{K}/n\overset{d}{\rightarrow }\mathbf{S}_{p}\text{ as }n\rightarrow
\infty .  \label{Equ3}
\end{equation}
Applying the strong law of large numbers (conditionally given $\mathbf{S}%
_{p} $), the above convergence in law also holds almost surely.\newline

With the main results being from (\ref{f9}-\ref{f10b1}), the present study
can perhaps be summarized for the different regimes as follows: 
\begin{equation*}
\begin{array}{llllll}
\backslash & 
\begin{array}{l}
\text{Range of the} \\ 
\text{parameters}
\end{array}
& Q\overset{d}{\sim } & \mathbf{E}\left( Q\right) \sim & K\left( 1\right) 
\overset{d}{\sim } & \mathbf{K}\overset{d}{\sim } \\ 
\mathbf{1} & 
\begin{array}{l}
n,p\rightarrow \infty \text{, }\frac{n}{p}=\mu ^{*} \\ 
\nu \text{ fixed}
\end{array}
& \text{(\ref{equ1})} & 
\begin{array}{l}
n\frac{1-\nu ^{\theta ^{*}}}{\mu ^{*}} \\ 
\theta ^{*}=\frac{\mu ^{*}\nu }{1-\nu }
\end{array}
& 
\begin{array}{l}
\text{negative } \\ 
\text{binomial}
\end{array}
& 
\begin{array}{l}
\text{multivariate} \\ 
\text{neg. binomial}
\end{array}
\\ 
\mathbf{2} & 
\begin{array}{l}
n,p\rightarrow \infty \text{, }\frac{n}{p}\rightarrow 0 \\ 
\nu \text{ fixed}
\end{array}
& 
\begin{array}{l}
\text{(\ref{equ1})} \\ 
\gamma =\frac{n\nu }{1-\nu }
\end{array}
& n\frac{\nu \log \left( 1/\nu \right) }{1-\nu } & 
\begin{array}{l}
\mid K\left( 1\right) \geq 1: \\ 
\log \text{-series}
\end{array}
& \text{Ewens }\gamma =\frac{n\nu }{1-\nu } \\ 
\mathbf{3} & 
\begin{array}{l}
n\rightarrow \infty \text{, }\nu \rightarrow 0 \\ 
\nu n=\lambda \text{, }\nu p\rightarrow \infty
\end{array}
& 
\begin{array}{l}
\text{(\ref{equ1})} \\ 
\gamma \sim \lambda
\end{array}
& \lambda \log n & \text{(\ref{f3})} & \text{Ewens }\gamma \sim \lambda \\ 
\mathbf{4} & 
\begin{array}{l}
n\rightarrow \infty \text{, }\nu \rightarrow 0 \\ 
\nu n\log n=c
\end{array}
& 1+\text{Poi}\left( c\right) & 1+c & \text{(\ref{f3})} & \text{(\ref{f8a})}
\\ 
\mathbf{5} & 
\begin{array}{l}
n\rightarrow \infty \text{, }\nu \rightarrow 0 \\ 
p\text{ fixed, }n\nu =\lambda
\end{array}
& \text{(\ref{f10a})}\emph{\ } & \text{(\ref{f10b2})} & 
\begin{array}{l}
n\text{beta}\left( \theta ^{*},\lambda -\theta ^{*}\right) \\ 
\theta ^{*}=\lambda /p
\end{array}
& n\mathbf{S}_{p},\text{ (\ref{Equ3})}
\end{array}
\end{equation*}
\newline

\textbf{Acknowledgments:} The author acknowledges partial support from the
labex MME-DII (\emph{Mod\`{e}les Math\'{e}matiques et \'{E}conomiques de la
Dynamique, de l' Incertitude et des Interactions}), ANR11-LBX-0023-01. This
work also benefited from the support of the Chair ``\emph{Mod\'{e}lisation
Math\'{e}matique et Biodiversit\'{e}}'' of Veolia-Ecole
Polytechnique-MNHN-Fondation X.

\end{document}